\documentclass[11pt]{article}
\usepackage[left=1in,top=1in,right=1in,bottom=1in,head=.1in,nofoot]{geometry}

\usepackage{amsmath}
\usepackage{color}
\usepackage{multirow}
\usepackage{graphicx}
\usepackage{multirow}
\usepackage{amssymb, amsthm}
\usepackage{url}
\usepackage{setspace}

\usepackage{times}
\usepackage{bm}
\usepackage{natbib}

\usepackage[plain,noend]{algorithm2e}

\theoremstyle{definition}
\newtheorem{lemma}{Lemma}
\newtheorem{theorem}{Theorem}

\newtheorem{corollary}{Corollary}

\newtheorem{assumption}{Assumption}

\newtheorem{condition}{Condition}
\newtheorem{proposition}{Proposition}

\newcommand{\yl}{Y_{i,t+1}}
\newcommand{\yo}{Y_{it}}

\newcommand{\yll}{Y_{i,t+1}(1)}

\newcommand{\ylo}{Y_{i,t+1}(0)}
\newcommand{\yoo}{Y_{it}(0)}

\newcommand{\di}{G_i}

\newcommand{\be}{\begin{equation}}
\newcommand{\en}{\end{equation}}
\newcommand{\bea}{\begin{eqnarray}}
\newcommand{\ena}{\end{eqnarray}}
\newcommand{\ba}{\begin{array}}

\newcommand{\ea}{\end{array}}

\newcommand{\sumi}{\sum_{i=1}^{n}}

\newcommand{\pr}{\textup{pr}}

\def\ci{\perp\!\!\!\perp}
\newcommand\bE{E} % expectation

\newcommand{\att}{ {\mathrm{ATT}}}

\newcommand{\dd}{\mathrm{d}}

\newcommand{\DID}{ {\mathrm{DID}}}

\newcommand{\IG}{ {\mathrm{LDV}}}

\newcommand*{\T }{{\mkern-1.5mu\mathsf{T}}}

\makeatletter
\renewcommand{\algocf@captiontext}[2]{#1\algocf@typo. \AlCapFnt{}#2} % text of caption
\def\@algocf@capt@plain{top}
\renewcommand{\algocf@makecaption}[2]{%
  \addtolength{\hsize}{\algomargin}%
  \sbox\@tempboxa{\algocf@captiontext{#1}{#2}}%
  \ifdim\wd\@tempboxa >\hsize%     % if caption is longer than a line
    \hskip .5\algomargin%
    \parbox[t]{\hsize}{\algocf@captiontext{#1}{#2}}% then caption is not centered
  \else%
    \global\@minipagefalse%
    \hbox to\hsize{\box\@tempboxa}% else caption is centered
  \fi%
  \addtolength{\hsize}{-\algomargin}%
}
\makeatother

\begin{document}

\title{A bracketing relationship between difference-in-differences and lagged-dependent-variable adjustment}
\author{Peng Ding
\footnote{Department of Statistics, University of California, 425 Evans Hall, Berkeley, California
94720, U.S.A. Email: pengdingpku@berkeley.edu}
~and Fan Li
\footnote{Department of Statistical Science, Duke University, Box 90251, Durham, North Carolina 27708 U.S.A. Email: fl35@duke.edu}
}
\date{}
\maketitle

\begin{abstract}
Difference-in-differences is a widely-used evaluation strategy that draws causal inference from observational panel data. Its causal identification relies on the assumption of parallel trends, which is scale dependent and may be questionable in some applications. A common alternative is a regression model that adjusts for the lagged dependent variable, which rests on the assumption of ignorability conditional on past outcomes. In the context of linear models, \citet{APbook} show that the difference-in-differences and lagged-dependent-variable regression estimates have a bracketing relationship. Namely, for a true positive effect, if ignorability is correct, then mistakenly assuming parallel trends will overestimate the effect; in contrast, if the parallel trends assumption is correct, then mistakenly assuming ignorability will underestimate the effect. We show that the same bracketing relationship holds in general nonparametric (model-free) settings. We also extend the result to semiparametric estimation based on inverse probability weighting. We provide three examples to illustrate the theoretical results with replication files in \citet{ding2019bracketingData}.
\end{abstract}


\section{Introduction}
Difference-in-differences is a popular evaluation strategy in the social sciences; it makes causal comparisons from observational panel data by exploiting variation across time \citep{ashenfelter1978estimating,bertrand2004much, APbook, bechtel2011lasting, keele2013much, malesky2014impact, keele2016patterns, callaway2019difference}. The key assumption underlying difference-in-differences is \emph{parallel trends}, that is, the counterfactual trend behavior of treatment and control groups, in the absence of treatment, is the same, possibly conditioning on some observed covariates \citep{Heckman1997, Abadie2005}. In practice, the parallel trends assumption can be questionable because unobserved confounders may have time-varying effects on the outcomes. A common alternative method is a regression model that adjusts for the lagged dependent variables  \citep{ashenfelter1978estimating}, which assumes \emph{ignorability} conditional on past outcomes and observed covariates.

Difference-in-differences and lagged-dependent-variable adjustment---also known respectively as the gain score estimator and the analysis of covariance estimator in  sociology and psychology---are two different methods relying on different identification assumptions. Extensive conceptual, empirical and numerical comparisons between the two methods have been made in the literature \citep[e.g.,][]{allison1990change, maris1998covariance, van2013ancova, ryan2015we, o2016estimating}. In particular, in the context of linear models,  \cite{APbook} show that difference-in-differences and lagged-dependent-variable regression estimators have a \emph{bracketing} relationship. Namely, for a true positive effect, if the ignorability assumption is correct, then mistakenly assuming parallel trends will overestimate the effect; in contrast, if the parallel trends assumption is correct, then mistakenly assuming ignorability will underestimate the effect. The opposite holds for a true negative effect.

The bracketing relationship is important in practice. Though we usually do not know which one of the two assumptions is true in real applications, we can analyze the data under each assumption and treat the estimates as the upper and lower bounds of the true effect. However, the linear setting in \cite{APbook} is restrictive, particularly for applications with non-continuous outcomes. For example, binary outcomes are common in political science \citep[e.g.,][]{keele2013much, malesky2014impact} and health studies \citep[e.g.,][]{stuart2014using}; count outcome are common in transportation safety studies where the before-after design is popular \citep[e.g.,][]{hauer1997observational}. Moreover, the parallel trends assumption is functional-form dependent \citep{Athey2006}. Therefore, an extension to nonlinear settings is relevant for both theory and practice. In this paper, we prove that, within the canonical two-period two-group setting, the same bracketing relationship holds in general nonparametric and semiparametric settings. We give three examples to illustrate the theoretical results.

\section{Setup}
\subsection{Difference-in-differences}
We proceed under the potential outcomes framework \citep{Neyman:1923, rubin1974estimating}. We consider the basic two-period two-group panel design, where a sample of units, indexed by $i\in \{1,...,n\}$, are drawn from a target population of two groups, labeled by $G_i =0$ or $1$.  Each unit can potentially be assigned to a treatment $d$, with $d = 1$ for the active treatment and $d = 0$ for the control. Units in both groups are followed in two periods of time $T$, with $T = t$ and $T   = t+1$ denoting the before and after period, respectively.  The treatment is only administered to the group with $G_i=1$ in the after period. For each unit $i$, let $D_{iT}$ be the observed treatment status at time $T$. The above design implies $D_{it}=0$ for all units and $D_{i,t+1}=1$ for the units in group $G_i=1$; thus $G_i=D_{i,t+1}$. Assume that each unit has two potential outcomes in each period, $\{ Y_{iT}(1) , Y_{iT}(0) \}$ for $T=t$ and $t+1$, and only the one corresponding to the observed treatment status, $Y_{iT}=Y_{iT}(D_{iT})$, is observed.  Therefore, $\yo=\yoo$ and $\yl=(1-\di)\ylo+\di\yll$. For each unit, a vector of pre-treatment covariates $X_i$ are also observed in the before period.

In the two-period two-group panel design, the target estimand is usually the average treatment effect for the treated (ATT) \citep{Abadie2005, APbook, lechner2011estimation}:
\begin{equation} \label{est:att}
\tau_{\att}\equiv\bE\{\yll-\ylo\mid G_i=1\}=\mu_1-\mu_0,
\end{equation}
where $\mu_1= \bE\{  Y_{i,t+1}(1)  \mid G_i=1\}$ and  $\mu_0= \bE\{  Y_{i,t+1}(0)  \mid G_i=1\}$.
When the outcome is discrete, ratio versions of $\tau_{\att}$ are often of interest, such as
\begin{equation} \label{est:rr}
\gamma_\att \equiv {\bE\{  \yll \mid G_i=1\}  }/{\bE\{  \ylo\mid G_i=1\} }=\mu_1/\mu_0,
\end{equation}
which is the causal risk ratio for binary outcomes and the causal rate ratio for count outcomes.

The quantity  $\mu_1$ equals $\bE(Y_{i,t+1}\mid G_i=1)$, and thus is directly estimable from the observed data, e.g., by the moment estimator $\bar{Y}_{1,t+1} =   \sumi G_iY_{i,t+1}/\sumi G_i$. In contrast, the quantity $\mu_0$, the counterfactual outcome for the treatment group in the after period in the absence of treatment, is not observable and must rely on additional assumptions to identify. The central task in this design is to use the observed data to estimate the counterfactual $\mu_0$. Any consistent estimator of $\mu_0$ leads to consistent estimators of $\tau_{\att}$ and $\gamma_{\att}$.

With difference-in-differences, the key for identifying $\mu_0$ is the \emph{parallel trends} assumption.
\begin{assumption}[Parallel trends] \label{as:para}
$\bE\{\ylo-\yoo\mid X_i, G_i=1\}=\bE\{\ylo-\yoo\mid X_i, G_i=0\}.$
\end{assumption}

The parallel trends assumption requires that, conditional on covariates $X_i$, the average outcomes in the treated and control groups in the absence of treatment would have followed parallel paths over time. Under Assumption \ref{as:para}, we have the non-parametric identification formula for $\mu_{0}$:
\begin{eqnarray}
\tilde{\mu}_{0,\DID}
&=& E\big[    E\{ Y_{it}(0)\mid X_i, G_i=1  \} + E\{   Y_{i,t+1}(0) - Y_{it}(0) \mid X_i, G_i=0   \} \mid G_i=1\big] \nonumber \\
&=& \bE(Y_{it}\mid G_i=1) + \bE\{ \bE( Y_{i,t+1} - Y_{it} \mid X_i, G_i=0  )  \mid G_i=1\} \nonumber \\
&=&  \bE(Y_{it}\mid G_i=1) +  \int  \bE( Y_{i,t+1} - Y_{it} \mid X_i = x, G_i=0  )   F_{X|G=1}(\text{d} x),
\label{eq:identify_x}
\end{eqnarray}
where $F_{X|G=1}(  x) = \text{pr}(X\leq x \mid G=1)$ is the distribution of $X$ in the treatment group. 
All terms of the right hand side of \eqref{eq:identify_x} are identifiable from the observed data. A stronger version of Assumption \ref{as:para} imposes  parallel trends without conditioning on covariates, under which we can write
\begin{eqnarray}
\tilde{\mu}_{0,\DID}&=&   \bE(Y_{it} \mid  G_i=1 )+  \bE(Y_{i,t+1}\mid G_i=0)- \bE(Y_{it} \mid  G_i=0 ) .
\label{eq:identify}
\end{eqnarray}
Based on the identification formula \eqref{eq:identify}, a moment estimator of $\tau_{\att}$ is
\begin{eqnarray}\label{eq::did_mom}
\hat{\tau}_{\DID} = (\bar{Y}_{1,t+1}-\bar{Y}_{1,t}) -(\bar{Y}_{0,t+1}-\bar{Y}_{0,t}),
\end{eqnarray}
where $\bar{Y}_{g,T}$ is the mean observed outcome for group $g$ at time $T$ $(g=0,1; T=t,t+1)$. The form of this estimator underlies the name ``difference-in-differences''.

A well-known limitation of the difference-in-differences approach is that the parallel trends assumption depends on the scale of the outcome \citep{Athey2006, lechner2011estimation}. Specifically, the parallel trends assumption may hold for the original $Y$ but not for a nonlinear monotone transformation of $Y$, for example, $\log Y$. This scale-dependence restricts the use of difference-in-differences in settings with non-Gaussian and discrete outcomes.

\subsection{Lagged-dependent-variable adjustment}
In the treatment-control panel design, a class of alternative methods rely on the assumption of ignorability conditional on the lagged dependent variable, that is, in the absence of treatment, the outcomes for the treated and control groups would have the same distributions, conditional on their lagged outcome and covariates.
\begin{assumption}[Ignorability]\label{assume:ig}
$Y_{i,t+1} (0) \ci  G_i \mid (Y_{it}, X_i)$.
\end{assumption}

Under ignorability, we have the following nonparametric identification formula of $\mu_{0}$:
\begin{eqnarray}
  \tilde{\mu}_{0,\IG} 
&=&   \bE\big[   \bE\{  Y_{i,t+1}(0)\mid G_i=1, Y_{it}, X_i\}  \mid G_i=1  \big]  \nonumber  \\
  &=& \bE\{   \bE(Y_{i,t+1}\mid G_i=0, Y_{it}, X_i)  \mid G_i=1  \}  \nonumber  \\
&=& \int   \bE(Y_{i,t+1}\mid G_i=0, Y_{it} = y, X_i =x)  F_{Y_t,X|G=1}( \text{d}  y, \text{d} x), \label{eq::ldv-identificationwithx}
\end{eqnarray}
where $F_{Y_t,X|G=1}(   y,   x)$ is the joint distribution of $(Y_t,X)$ in the treatment group.
The form of $\tilde{\mu}_{0,\IG}$ is identical to the traditional identification formula for the average treatment effect for the treated in observational cross-sectional studies. We can specify a model for $\bE\{Y_{i,t+1}(0)\mid Y_{it}, G_i, X_i\}$, based on which we impute the counterfactual mean $ \mu_0 = \bE\{Y_{i,t+1}(0)\mid G_i=1\}$ by averaging over $Y_t$ and $X$ and thus obtain a consistent estimator for $\tau_{\att}$.

In contrast to the parallel trends assumption, the ignorability assumption is scale free. Three popular methods under the ignorability assumption are the synthetic control method \citep{abadie2003economic, abadie2015comparative}, matching \citep{Heckman1997} or regression adjustment \citep{ashenfelter1978estimating} of the lagged dependent variable. Among these, the lagged-dependent-variable adjustment approach is the easiest to implement. Through extensive simulations, \cite{o2016estimating} have found that, when the parallel trends assumption does not hold, the lagged-dependent-variable regression adjustment approach produces the most efficient and least biased estimates among these three methods.

\section{Theory} \label{sec:theory}

Our goal in this section is to establish the analytical relationship between the difference-in-difference and lagged-dependent-variable adjustment estimators under general settings. For notational simplicity, we condition on the covariates $X$ and thus ignore them in the discussion.

\subsection{Bracketing relationship in linear models}\label{sec::linear}
We start with the simple case of linear regressions. Specifically, the difference-in-differences approach is usually implemented via a linear fixed-effects model:
\begin{eqnarray}\label{eq::lfem}
\bE(   Y_{iT} \mid D_{iT}, X_i )  =\alpha_i  + \lambda_T +  \tau D_{iT}  ,\quad (i=1,\ldots,n; T=t,t+1)
\end{eqnarray}
where $\alpha_i$ is the individual fixed effect and $\lambda_T$ is the time-specific fixed effect. When model \eqref{eq::lfem} is correct, the coefficient $\tau$ equals the estimand $\tau_{\att}$; any consistent estimator of $\tau$ in \eqref{eq::lfem} is also consistent for $\tau_{\att}$. By taking the difference between outcomes at time points $t$ and $t+1$ in \eqref{eq::lfem}, we can eliminate the individual fixed effects $\alpha_i$. Because $G_i =   D_{i,t+1} - D_{it}$, we have
$
\bE(   Y_{i,t+1} - Y_{it} \mid G_i )  =  (\lambda_{t+1} - \lambda_t) +  \tau G_i .
$
Therefore, we can fit a linear regression of the difference $Y_{i,t+1} - Y_{it} $ on the group indicator $G_i$ to estimate $\tau$. The resulting ordinary least squares estimator is the difference between the sample means of $Y_{i,t+1} - Y_{it} $ in the treated and control groups, and thus it equals $\hat{\tau}_\DID$ defined in \eqref{eq::did_mom}.

The lagged-dependent-variable adjustment method can be implemented via linear models in two ways. In the first approach, motivated by \eqref{eq::ldv-identificationwithx}, we can fit an ordinary least squares line $\hat{\bE}(Y_{t+1}\mid G=0, Y_t = y)  =   \hat{\alpha} + \hat{\beta} Y_t$ using only the control units; then we obtain $ \hat{\mu}_{0,\IG} =  \hat{\alpha} + \hat{\beta} \bar{Y}_{1,t}$ as the sample analog of $\tilde{\mu}_{0,\IG}$ and $\hat{\tau}_{\IG} = \bar{Y}_{1,t+1} - \hat{\mu}_{0,\IG} $ as the estimate of $\tau_\att$. In the second approach, as in  \citet[][Chapter 5.4]{APbook}, we can use the following linear model:
\begin{equation}
 \label{eq:LDV}
\bE(Y_{i,t+1}\mid  Y_{it}, G_i)= \alpha + \beta  Y_{it} + \tau G_i   .
\end{equation}
When model \eqref{eq:LDV} is correct, the coefficient $\tau$ equals the causal estimand $\tau_{\att}$, and any consistent estimator of $\tau$ is consistent for $\tau_{\att}$.
We can fit the ordinary least squares line $\hat{\bE}(Y_{t+1}\mid G,Y_t) = \hat{\alpha}  +  \hat{\tau}_{\IG}' G + \hat{\beta}' Y_t $ using all units and take the coefficient $\hat{\tau}_{\IG}'$ as an estimate of $\tau_{\att}$.
We have the following expressions for the two estimators $\hat{\tau}_{\IG} $ and $\hat{\tau}_{\IG}'$ (the proof is given in the appendix).

\begin{proposition}\label{prop1}
Without covariates, the two lagged-dependent-variable adjustment estimates are
\begin{eqnarray}
\label{eq::ldv-linear}
\hat{\tau}_{\IG} = (\bar{Y}_{1,t+1} - \bar{Y}_{0,t+1} ) - \hat{\beta}  (\bar{Y}_{1,t} - \bar{Y}_{0,t} ),\quad
\hat{\tau}_{\IG} ' = (\bar{Y}_{1,t+1} - \bar{Y}_{0,t+1} ) - \hat{\beta}'  (\bar{Y}_{1,t} - \bar{Y}_{0,t} ).
\end{eqnarray}
\end{proposition}

These two estimates in \eqref{eq::ldv-linear} differ from the moment difference-in-differences estimate $\hat{\tau}_{\DID} = (\bar{Y}_{1,t+1} - \bar{Y}_{0,t+1} ) -  (\bar{Y}_{1,t} - \bar{Y}_{0,t} ) $ only in the coefficients $\hat{\beta}$ and $\hat{\beta}'.$ Consider the case with $\hat{\beta}$ or $\hat{\beta}'$ larger than $0$ but smaller than $1$. The sign of $\hat{\tau}_{\DID} - \hat{\tau}_{\IG} $ or $\hat{\tau}_{\DID} - \hat{\tau}_{\IG}' $ depends on the sign of $\bar{Y}_{1,t} - \bar{Y}_{0,t}$. If the treatment group has larger lagged outcome $Y_t$ on average, then $\hat{\tau}_{\DID} < \hat{\tau}_{\IG} $; if the treatment group has smaller $Y_t$ on average, then $\hat{\tau}_{\DID} > \hat{\tau}_{\IG} $. In the special case with $\hat{\beta} = 1$ or $\hat{\beta}' = 1$, they are \emph{identical}: $\hat{\tau}_{\DID} = \hat{\tau}_{\IG}$ or $\hat{\tau}_{\DID}' = \hat{\tau}_{\IG}$. How much $\hat{\beta}$ or $\hat{\beta}'$ deviates from 1 indicates how different the two estimates are. We will see this phenomenon in the examples in Section \ref{sec:examples}. Importantly, the discussion in this subsection holds without imposing any stochastic assumptions. That is, Proposition \ref{prop1} is a purely numerical result. 
 In contrast, the bracketing relationship in \citet[][Chapter 5.4]{APbook} is proven under the linear model assumptions.

\citet{gelman2007did} pointed out that restricting $\beta=1$ in \eqref{eq:LDV} gives identical least squares estimators for $\tau$ from models \eqref{eq::lfem} and \eqref{eq:LDV}, which is also evident from Proposition \ref{prop1}. However, the nonparametric identification Assumptions \ref{as:para} and \ref{assume:ig} are not nested, and the difference-in-differences estimator is not a special case of the lagged-dependent-variable adjustment estimator in general. Therefore, it is natural to investigate whether \citet{APbook}'s result is unique to the linear models \eqref{eq::lfem} and \eqref{eq:LDV}. In the next subsection, we generalize the bracketing relationship to model-free settings.

\subsection{Nonparametric bracketing relationship} \label{sec:nonpara}

For notational simplicity, below we also drop the subscript $i$. Under ignorability, the nonparametric identification formula \eqref{eq::ldv-identificationwithx} of $\mu_{0}$ simplifies to
\begin{equation}\label{eq:identifyIG}
  \tilde{\mu}_{0,\IG} =\bE\{   \bE(Y_{t+1}\mid G=0, Y_t)  \mid G=1  \}
  = \int \bE(Y_{t+1}\mid G=0, Y_t=y) F_{Y_t}(\dd y\mid G=1) ,
\end{equation}
where $F_{Y_t}(y\mid G=g) = \pr(Y_t\leq y\mid G=g)$ is the cumulative distribution function of $Y_t$ for units in group $g$ $(g=0,1)$. The form of $\tilde{\mu}_{0,\IG}$ is identical to the identification formula for the ATT estimand in cross-sectional studies. 

To compare $\tilde{\tau}_{\DID} $ and $\tilde{\tau}_{\IG}$ without imposing any functional form of the outcome model, we first obtain the following analytical difference between $\tilde{\mu}_{0,\DID}$ and $\tilde{\mu}_{0,\IG}$ (the proof is given in the appendix).
\begin{lemma}\label{lemma::diff}
The difference between $\tilde{\mu}_{0,\DID} $ and $\tilde{\mu}_{0,\IG}$ is
$$
\tilde{\mu}_{0,\IG}-\tilde{\mu}_{0,\DID} = \int \Delta(y) F_{Y_t}(\dd y\mid G=1) -  \int \Delta(y) F_{Y_t}(\dd y\mid G=0) ,
$$
where $  \Delta(y) = \bE(Y_{t+1}\mid G=0, Y_t = y) - y.$
\end{lemma}

The quantity $ \Delta(y) = \bE(Y_{t+1}\mid G=0, Y_t = y) - y = \bE(Y_{t+1} - Y_t\mid G=0, Y_t = y)$ equals the expectation of the change in the outcome conditioning on the lagged outcome in the control group.
Lemma \ref{lemma::diff} suggests that the relative magnitude of  $\tilde{\mu}_{0,\DID} $ and $\tilde{\mu}_{0,\IG}$ depends on (a) the expectation of the before-after difference $Y_{t+1} - Y_t$ conditional on $Y_t$ in the control group, and (b) the difference between the distribution of the before outcome $Y_t$ in the treated and control groups. Both are important characteristics of the underlying data generating process, which measures (a) the dependence of the outcome on the lagged outcome and (b) the dependence of the treatment assignment on the lagged outcome, respectively. In particular, if $Y_t\ci  G$ or equivalently $F_{Y_t}(y\mid G=1) = F_{Y_t}(y\mid G=0)$, then $\tilde{\mu}_{0,\IG} = \tilde{\mu}_{0,\DID}.$

To reach the main conclusion, we introduce two additional conditions regarding the quantities in Lemma \ref{lemma::diff}.
The first is a stationarity condition on the outcome.
\begin{condition}
[Stationarity]\label{assume::stationary}
$ \partial \bE(Y_{t+1}\mid G=0, Y_t = y) / \partial y < 1    $ for all $y.$
\end{condition}

In a linear model for $\bE(Y_{t+1}\mid G=0, Y_t = y)$, Condition \ref{assume::stationary} requires that, in the control group, the regression coefficient of the outcome $Y_{t+1}$ on the lagged outcome $Y_t$ is smaller than $1$; this is also invoked by \cite{APbook}. Its sample version is $\hat{\beta} <1$ or $\hat{\beta}' < 1$ as in Section \ref{sec::linear}. In general, Condition \ref{assume::stationary} ensures that the time series of the outcomes would not grow infinitely as time, which is reasonable in most applications.
%; indeed, this condition is also assumed in \cite{APbook}.

The second condition describes the treatment assignment mechanism with respect to the lagged outcome, with two opposite versions.
\begin{condition}
[Stochastic Monotonicity]\label{assume::monotone}
(a) $F_{Y_t}(y\mid G=1) \geq  F_{Y_t}(y\mid G=0)$ for all $y$; (b) $F_{Y_t}(y\mid G=1) \leq  F_{Y_t}(y\mid G=0)$ for all $y$.
\end{condition}

Condition \ref{assume::monotone}(a) implies that the treated group has smaller lagged outcome compared to the control group, and Condition \ref{assume::monotone}(b) implies the opposite relationship. In the case of linear models, Condition \ref{assume::monotone}(a) or (b) reduces to the \emph{average} lagged outcome in the treated group is smaller or larger than that in the control group, respectively.

Because they only involve observed variables, Conditions \ref{assume::stationary} and \ref{assume::monotone} are testable empirically. Specifically, to check Condition \ref{assume::stationary}, we can estimate the derivative of the conditional mean function $\bE(Y_{t+1}\mid G=0, Y_t = y)$; to check Condition \ref{assume::monotone}, we can visually compare the empirical cumulative probability distributions of the outcomes in the treatment and control groups. These conditions hold in many applications, e.g., in the examples in Section \ref{sec:examples}. In contrast, Assumption \ref{as:para} and \ref{assume:ig} are in general untestable.

Under Conditions \ref{assume::stationary} and \ref{assume::monotone}, we have the following results on the bracketing relationship on $\tilde{\tau}_{\DID} $ and $\tilde{\tau}_{\IG}$ in a nonparametric setting; see the appendix for the proof.
\begin{theorem}\label{thm::monoATT}
If Conditions \ref{assume::stationary} and \ref{assume::monotone}(a) hold, then $\tilde{\mu}_{0,\DID} \leq  \tilde{\mu}_{0,\IG}$, and thus $\tilde{\tau}_{\DID} \geq  \tilde{\tau}_{\IG}$ and $\tilde{\gamma}_{\DID} \geq  \tilde{\gamma}_{\IG}$; if Conditions \ref{assume::stationary} and \ref{assume::monotone}(b) hold, then $\tilde{\mu}_{0,\DID} \geq  \tilde{\mu}_{0,\IG}$, and thus $\tilde{\tau}_{\DID} \leq   \tilde{\tau}_{\IG}$ and $\tilde{\gamma}_{\DID} \leq  \tilde{\gamma}_{\IG}$.
\end{theorem}

Theorem \ref{thm::monoATT}  is a result on the relative magnitude between the two quantities $\tilde{\tau}_{\DID}$ and $\tilde{\tau}_{\IG}$ (and between $\tilde{\gamma}_{\DID} $ and $ \tilde{\gamma}_{\IG}$). On the one hand, Theorem \ref{thm::monoATT}  holds without requiring either Assumption \ref{as:para} or \ref{assume:ig}.  Specifically, under Stationarity and Stochastic Monotonicity (a), $\tilde{\tau}_{\DID} $ is larger than or equal to $  \tilde{\tau}_{\IG}$. Both of them can be biased for the true causal effect $\tau_{\att}$: if $ \tilde{\tau}_{\DID}  \geq  \tilde{\tau}_{\IG} \geq \tau_{\att}$, then $ \tilde{\tau}_{\DID}  $ over-estimates $\tau_{\att}$ more than $ \tilde{\tau}_{\IG} $; if $\tau_{\att}\geq  \tilde{\tau}_{\DID}  \geq  \tilde{\tau}_{\IG} $, then $\tilde{\tau}_{\IG} $ under-estimate $\tau_{\att}$ more than $\tilde{\tau}_{\DID} $; if $ \tilde{\tau}_{\DID}  \geq \tau_{\att}\geq   \tilde{\tau}_{\IG}  $, then $ \tilde{\tau}_{\DID}$ and $ \tilde{\tau}_{\IG} $ are the upper and lower bounds on $\tau_{\att}$. Analogous arguments apply under Stationarity and Stochastic Monotonicity (b). On the other hand, only under Assumption \ref{as:para} or \ref{assume:ig}, the quantities $\tilde{\tau}_{\DID}$ and $\tilde{\tau}_{\IG}$ have the interpretation as the nonparametric identification formulas of the causal estimand $\tau_\att$. We stress that the bracket $(\tilde{\tau}_{\DID}, \tilde{\tau}_{\IG})$ provides bounds for the true effect $\tau_{\att}$ if either Assumption \ref{as:para} or \ref{assume:ig} holds; however, it does not answer the question about whether the true effect falls inside or, if outside, which side of the bracket when neither Assumption \ref{as:para}  nor  \ref{assume:ig}  holds. The relationship under such a scenario is dependent on the specific true data generating model.

For discrete outcomes, equation \eqref{eq:identifyIG} reduces to $\tilde{\mu}_{0,\IG}= \sum_{y} E(Y_{t+1} \mid G=0, Y_t = y) \pr(Y_t = y \mid G=1)$, and the stationary condition becomes $\bE(Y_{t+1}\mid G=0, Y_t = y+1)-\bE(Y_{t+1}\mid G=0, Y_t = y)<1$ for all values of $y$. For the case of binary outcome, the stationary condition always holds because $0\leq\bE(Y_{t+1}\mid G=0, Y_t = y)\leq 1$ for $y=0,1$. Therefore, we only need to check the sign of the empirical counterpart of $\pr(Y_t=0 \mid G=1)-\pr(Y_t=0 \mid G=0)$. Specifically, if $\pr(Y_t=0\mid G=1)\geq \pr(Y_t=0\mid G=0)$, then $\tilde{\tau}_{\DID} \geq  \tilde{\tau}_{\IG}$ and $\tilde{\gamma}_{\DID} \geq  \tilde{\gamma}_{\IG}$; if $\pr(Y_t=0 \mid G=1)\leq \pr(Y_t=0 \mid G=0)$, then $\tilde{\tau}_{\DID} \leq  \tilde{\tau}_{\IG}$ and $\tilde{\gamma}_{\DID} \leq  \tilde{\gamma}_{\IG}$.

\subsection{Semiparametric bracketing relationship} \label{sec:semipara}

Under the parallel trends Assumption \ref{as:para}, \cite{Abadie2005} proposed a semiparametric inverse probability weighting estimator for $\tau_{\att}$ based on the following identification formula of $\mu_0$:
\begin{equation}\label{eq:ipw_DID}
\tilde{\mu}'_{0,\DID} = \bE\left\{G Y_{t}+\frac{e (1-G)(Y_{t+1}-Y_{t})}{1-e }\right\}\Big/\pr(G=1),
\end{equation}
where the propensity score is defined as $e  = \pr(G=1 )$. \cite{Abadie2005}'s estimator based on $\tilde{\mu}_{0,\DID} $ shares the same form as the inverse probability weighting estimator for the ATT in the cross-sectional setting, but replaces the outcome in the treatment group by the before-after difference $Y_{t+1}-Y_{t}$.
Similarly, under Assumption \ref{assume:ig}, we can construct a semiparametric estimator based on
\begin{equation}\label{eq:ipw_IG}
\tilde{\mu}'_{0,\IG}   = \bE\left\{  \frac{e(Y_t)}{1-e(Y_t) }  (1-G)  Y_{t+1}   \right\}\Big/ \pr(G=1),
\end{equation}
where the propensity score is defined as $e(Y_t) = \pr(G=1\mid Y_t)$.

Because \eqref{eq:ipw_DID} and \eqref{eq:ipw_IG} are alternative identification formulas for $\mu_0$, we can show that $(\tilde{\mu}_{0,\DID}, \tilde{\mu}_{0,\IG}) = (\tilde{\mu}'_{0,\DID}, \tilde{\mu}'_{0,\IG})$ and thus have the following corollary of Theorem \ref{thm::monoATT}.

\begin{corollary}\label{thm::IPW}
Theorem \ref{thm::monoATT} holds if $(\tilde{\mu}_{0,\DID}, \tilde{\mu}_{0,\IG})$ are replaced by $(\tilde{\mu}'_{0,\DID}, \tilde{\mu}'_{0,\IG})$.
\end{corollary}

Corollary \ref{thm::IPW} shows that the bracketing relationship between $\tilde{\tau}_{\DID}$ and $\tilde{\tau}_{\IG}$ does not depend on the forms of identification formulas and estimators.

\section{Examples}\label{sec:examples}
%Below we give three empirical examples to illustrate the theory in Section \ref{sec:theory}.

\subsection{Minimum wages and employment}
We re-analyze part of the classic \cite{card1994} study on the effect of a minimum wage increase on employment. Data were collected on the employment information at fast food restaurants in New Jersey and Pennsylvania before and after a minimum wage increase in New Jersey in 1992. The outcome is the number of full-time-equivalent employees at each restaurant.
%We do not use covariates. Table 3 of \cite{card1994} gives summary statistics of the data.
%Using the ordinary least squares of the first difference $Y_{t+1}-Y_t$ on the treatment indicator, we obtain
%. Using the ordinary least squares of the outcome $Y_{t+1}$ on the treatment indicator and $Y_t$, we obtain the lagged-dependent-variable regression estimate $\hat{\tau}_{\IG}=0.865$, and the coefficient of the lag outcome $\hat{\beta} = 0.475<1$.

\begin{figure}
\centering
\includegraphics[width=0.49\textwidth]{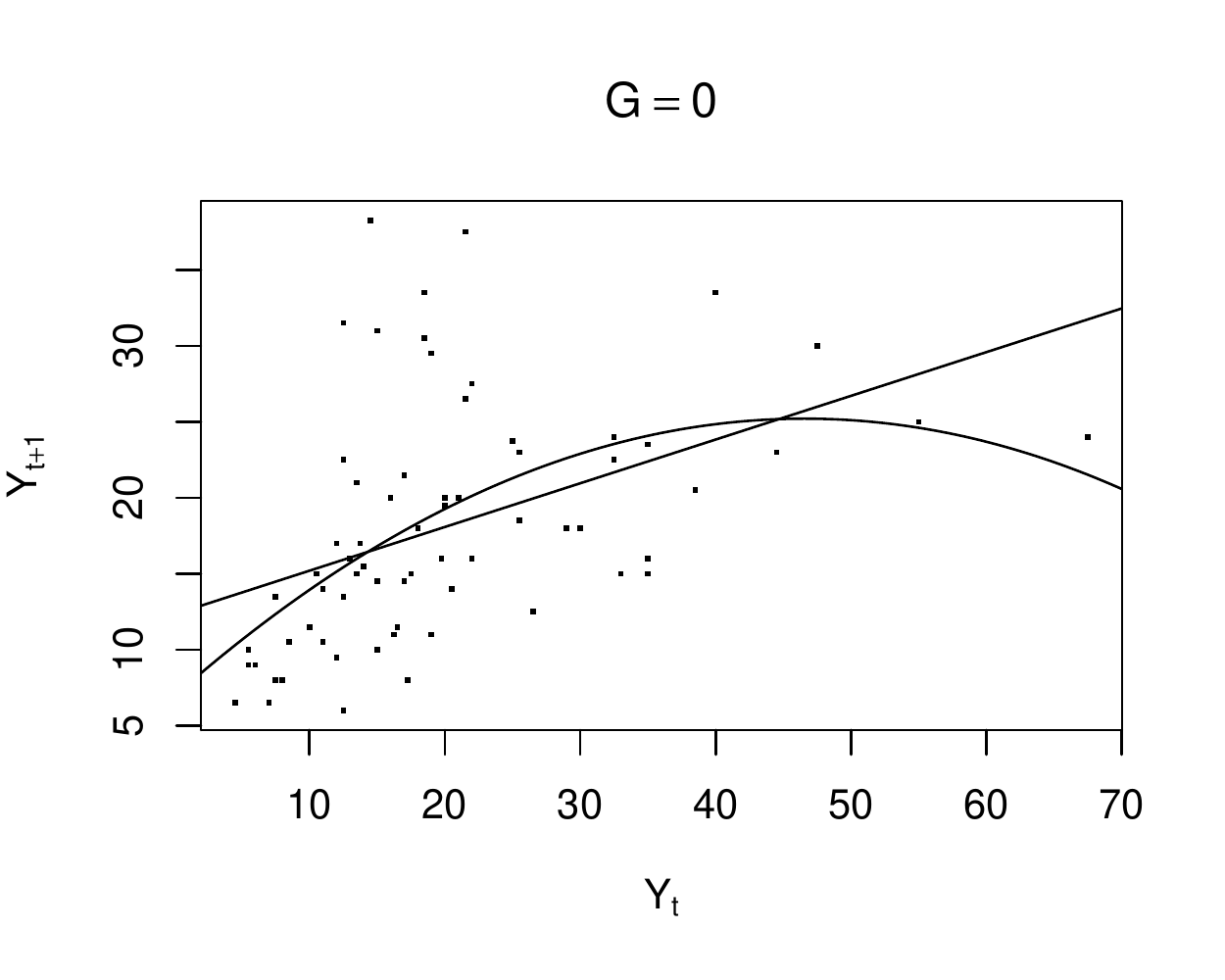}
\includegraphics[width=0.49\textwidth]{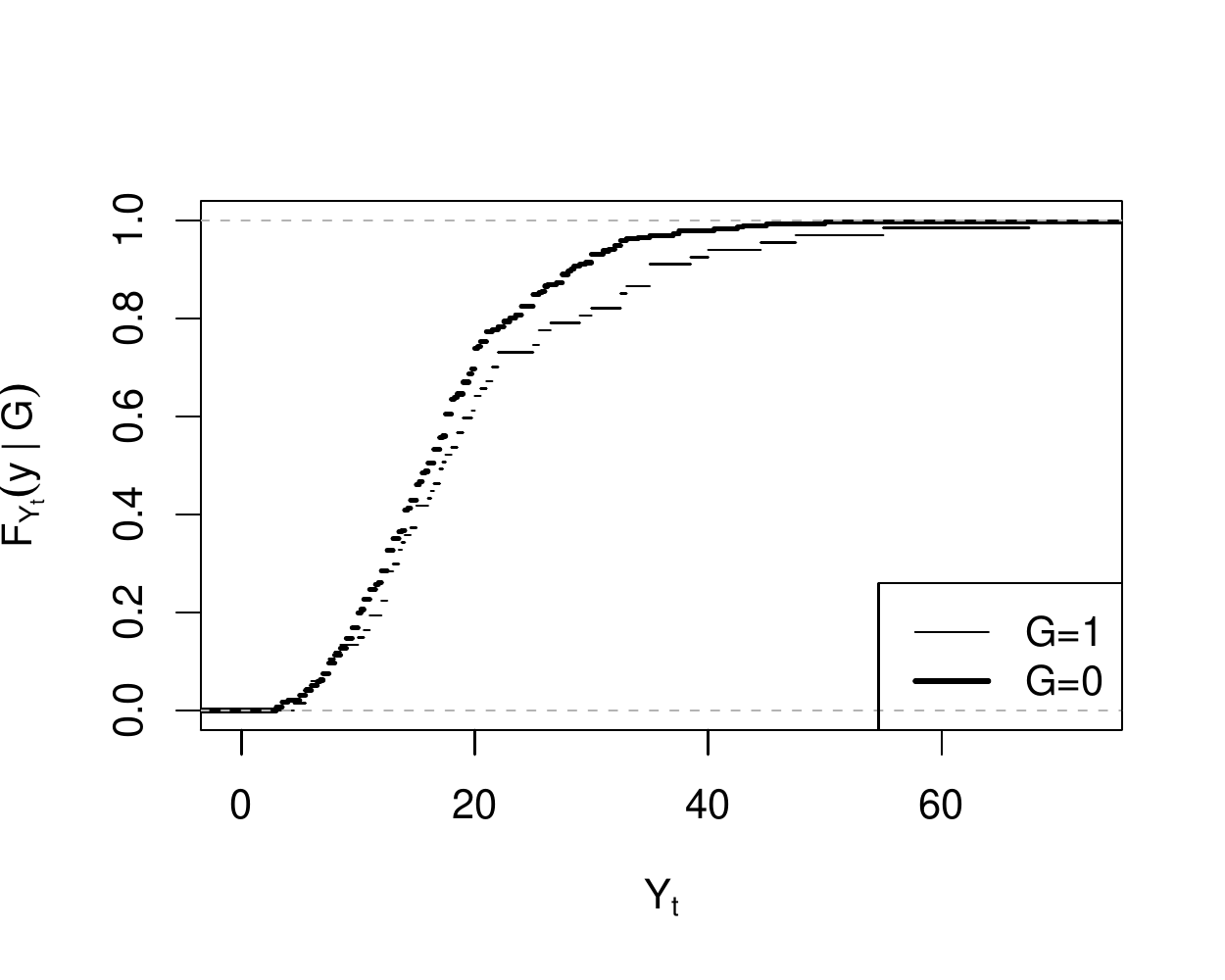}
\caption{\citet{card1994} study. Left: linear and quadratic fitted lines of $E(Y_{t+1}\mid G=0, Y_t)$. Right: empirical distribution functions $F_{Y_t}(y\mid G=g)\ (g=0,1)$ satisfy Stochastic Monotonicity.}\label{fig::card}
\end{figure}

The difference-in-differences estimate is $\hat{\tau}_{\DID}=2.446$, and the the lagged-dependent-variable adjustment estimates are $\hat{\tau}_{\IG}=0.302$ and $\hat{\tau}_{\IG}'=0.865$ with coefficients of the lag outcome $\hat{\beta} = 0.288 < 1$ and $\hat{\beta} '= 0.475<1$.
Meanwhile, because the sample means satisfy $ \bar{Y}_{1,t} - \bar{Y}_{0,t}  = 17.289 - 20.299<0$, our theoretical result predicts that $\hat{\tau}_{\DID} > \hat{\tau}_{\IG} (\text{or } \hat{\tau}_{\IG} ' )$, which exactly matches the relative magnitude of the empirical estimates above. In addition, if we adopt a quadratic specification of $E(Y_{t+1}\mid G=0, Y_t)$, the lagged-dependent-variable regression estimate becomes $\hat{\tau}_{\IG} = 0.275$, which is also smaller than $\hat{\tau}_{\DID}$. This is again coherent with our theory because Stationarity and Stochastic Monotonicity hold, depicted in Figure \ref{fig::card}. In this example, the differences between  $\hat{\tau}_{\DID} $ and $ \hat{\tau}_{\IG} ( \text{or } \hat{\tau}_{\IG} ' )$ are significant at level $0.05.$

\subsection{Electoral returns to beneficial policy}

We re-analyze the \citet{bechtel2011lasting} study on electoral returns to beneficial policy. We focus on the short-term electoral returns by analyzing the causal effect of disaster relief aid due to the 2002 Elbe flooding in Germany. The before period is 1998 and the after period is 2002. The units of analysis are electoral districts, the treatment is the indicator whether a district is affected by the flood, and the outcome is the vote share that the Social Democratic Party attains in that district.

\begin{figure}
\centering
\includegraphics[width=0.49\textwidth]{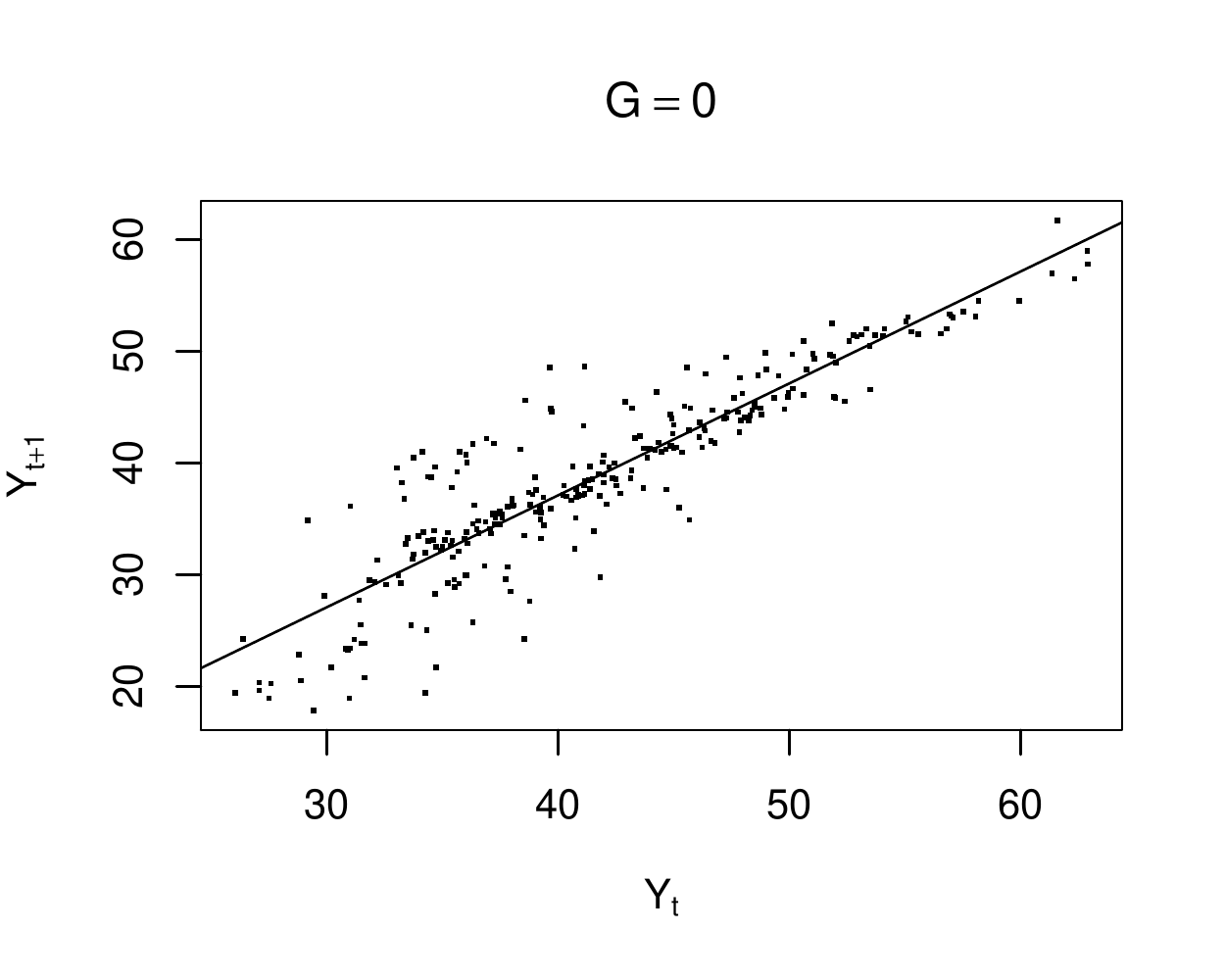}
\includegraphics[width=0.49\textwidth]{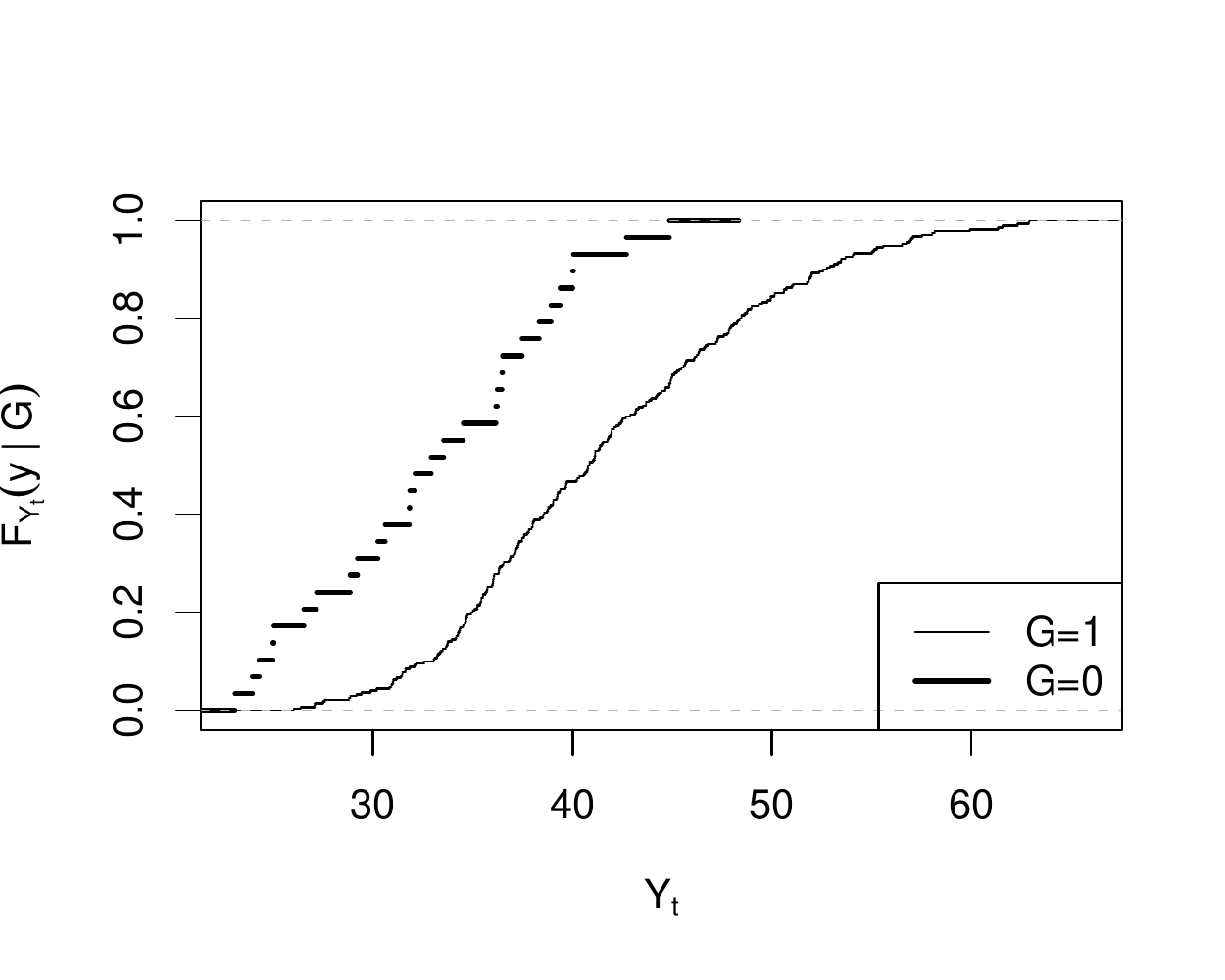}
\caption{\citet{bechtel2011lasting} study. Left: linear fitted lines of $E(Y_{t+1}\mid G=0, Y_t)$. Right: empirical distribution functions $F_{Y_t}(y\mid G=g)\ (g=0,1)$ satisfy Stochastic Monotonicity.}\label{fig::bh}
\end{figure}

The difference-in-differences estimate is $\hat{\tau}_{\DID} = 7.144$, and the lagged-outcome adjustment estimates are $\hat{\tau}_{\IG} =7.160$ and  $\hat{\tau}_{\IG} ' =7.121$ with coefficients of the lag outcome $\hat{\beta} = 1.002>1$ and $\hat{\beta}' = 0.997<1$. The relative magnitudes match our theory in Section \ref{sec::linear}. However, these estimates are almost identical because the coefficients of $Y_t$ are extremely close to $1$.  In this example, even though the empirical distributions of $F_{Y_t}(y\mid G=1)$ and $F_{Y_t}(y\mid G=0)$ differ significantly as Figure \ref{fig::bh} shows, the analysis is not sensitive to the choice between the difference-in-differences and lagged-dependent-variable adjustment estimates.

\subsection{A traffic safety intervention on crashes}
Outside the political science literature, the before-after treatment-control design is the state-of-art method in traffic safety evaluations \citep{hauer1997observational}, where count outcomes are common. Here we provide an example of evaluating the effects of rumble strips on vehicle crashes. Crash counts were collected on $n=1986$ road segments in Pennsylvania before (year 2008) and after (year 2012) the rumble strips were installed in 331 segments between year 2008 to 2012. The control group consists of 1655 sites matched to the treated sites on covariates including past accident counts, road characteristics, traffic volume. Table \ref{tab:crash} presents the crash counts classified by $Y_t$ and $Y_{t+1}$ for control and treatment groups, respectively.

 \begin{table}
  \centering
  \caption{Crash counts in the 1986 road sites in Pennsylvania (3+ means 3 or more crashes).}\label{tab:crash}
 \begin{tabular}{cc}
  (a) control group $G=0$ & (b) treated group $G=1$ \\
  \begin{tabular}{ccccccc}
  \hline
&& \multicolumn{4}{c}{$Y_{t+1}$}\\
&  &0 & 1 & 2 & 3+&total \\
\hline
\multirow{4}{*}{$Y_t$} & 0& 789 & 238 & 57 & 18&1102 \\
 & 1& 235 & 95 & 40 & 15 &385\\
 & 2& 61 & 37 & 11 &  6&115\\
 & 3+&  26 & 21 &  4 &  2& 53\\
 \hline
&total& 1111 & 391 & 112 &  41 & 1655\\
\hline
\end{tabular}
&
\begin{tabular}{ccccccc}
\hline
&& \multicolumn{4}{c}{$Y_{t+1}$}\\
&  &0 & 1 & 2 & 3+&total \\
\hline
\multirow{4}{*}{$Y_t$} &  0 &183 & 39  & 7 &  3 & 232\\
 & 1&  40 & 22 &  5 &  2& 69 \\
 & 2&  16 &  4 &  0 &  1 &  21 \\
 & 3+ &  2 &  6 &  0 &  1& 9 \\
 \hline
 &total& 241 & 71&  12&   7&   331\\
 \hline
\end{tabular}
\end{tabular}
\end{table}

We first examine the dichotomized outcome of whether there has been at least one crash in that site. As noted after Theorem \ref{thm::monoATT}, Condition \ref{assume::stationary} automatically holds for a binary outcome. We can verify that Condition \ref{assume::monotone}(a) holds because the empirical means suggest $\widehat{\pr}(Y_t=0\mid G=1)-\widehat{\pr}(Y_t=0\mid G=0)= 232/331 - 1102/1655 =.701-.666>0$. Therefore, applying Theorem \ref{thm::monoATT}, we predict that $\tilde{\tau}_{\DID}>\tilde{\tau}_{\IG}$ and $\tilde{\gamma}_{\DID}>\tilde{\gamma}_{\IG}$. Now we calculate the nonparametric estimate of $\mu_0$ under ignorability to be $\hat{\mu}_{0, \IG}=\sum_{y=0,1} \hat{E}(Y_{t+1} \mid G=0, Y_t = y) \widehat{\pr}(Y_t = y \mid G=1)=.324$, and under parallel trends to be $\hat{\mu}_{0, \DID}=.294$. Therefore, the empirical estimates suggest $\hat{\tau}_{\DID}>\hat{\tau}_{\IG}$ and $\hat{\gamma}_{\DID}>\hat{\gamma}_{\IG}$, which matches the theoretical prediction.

We then examine the original count outcome in Table \ref{tab:crash}. The sample means $\hat{\bE}(Y_{t+1}\mid G=0, Y_t = y)$ are $.374, .572, .670, .660$ for $y=0,1,2,3+$, respectively. Therefore, Condition \ref{assume::stationary} holds for all $y$. We can also verify that Condition \ref{assume::monotone}(a) holds because the sample probabilities are $\widehat{\pr}(Y_t\leq y\mid G=1)=.700, .909, .973$ and $\widehat{\pr}(Y_t\leq y\mid G=0)=.666, .898, .968$ for $y=0,1,2$, respectively. Therefore, applying Theorem \ref{thm::monoATT}, we predict that $\tilde{\tau}_{\DID}>\tilde{\tau}_{\IG}$ and $\tilde{\gamma}_{\DID}>\tilde{\gamma}_{\IG}$. Now we calculate the nonparametric estimate of $\mu_0$ under ignorability to be $\hat{\mu}_{0,\IG}=.438$, and under parallel trends to be $\hat{\mu}_{0,\DID}=.395$. Therefore, the empirical estimates suggest $\hat{\tau}_{\DID}>\hat{\tau}_{\IG}$ and $\hat{\gamma}_{\DID}>\hat{\gamma}_{\IG}$, which matches the theoretical prediction.

In this example, the differences between the $\hat{\gamma}_{\DID}$'s and the $\hat{\gamma}_{\IG}$'s are not significant at level $0.05.$

\section{Discussion}

We established a model-free bracketing relationship between the difference-in-differences and lagged-dependent-variable adjustment estimators in the canonical two-period two-group setting. In practice, we cannot validate the assumptions that justify these approaches. Therefore, a practical suggestion is to report results from both approaches and ideally to conduct sensitivity analyses allowing for violations of these assumptions.

Several directions are worth investigating. First, in the setting with $K+1$ time periods, we may consider a  model that incorporates both Model \eqref{eq::lfem} and \eqref{eq:LDV}: $E(Y_{i,T} \mid X_i, Y_{i,T-1}, G_i) = \alpha_i + \lambda_T + \beta Y_{i,T-1} + \tau G_i +  \theta^\T  X_i$ for $T=t+1, \ldots, t+K$. However, \citet{nickell1981biases} and \citet[][Section 5.3]{APbook} pointed out that identification and estimation under this model require much stronger assumptions. It is of interest to extend the bracketing relationship to this setting. Second, we focused on the average treatment effect on the treated; we can extend the result to other types (e.g. categorical and ordinal) of outcomes for which the averages are less interpretable \citep{lu2018treatment}.

\section*{Acknowledgments}

We are grateful to three reviewers, Jas Sekhon, Benjamin Lu, Olivia Angiuli, and Frank Li for helpful comments, to Avi Feller for kindly editing the paper, and to Eric Donnell for providing the transportation data. Peng Ding is supported by the National Science Foundation grant NSF-DMS1713152.

\appendix
\section*{Appendix: Proofs}

\noindent {\it Proof of Proposition \ref{prop1}.}
First, the ordinary least squares fit $\hat{\bE}(Y_{t+1}\mid G=0, Y_t = y)  =   \hat{\alpha} + \hat{\beta} Y_t$ using the control units must satisfy $\hat{\alpha} = \bar{Y}_{0,t+1} - \hat{\beta} \bar{Y}_{0,t}$. Therefore,
$$
\hat{\tau}_{\IG} = \bar{Y}_{1,t+1} - \hat{\mu}_{0,\IG}
= \bar{Y}_{1,t+1} - \hat{\alpha}  - \hat{\beta} \bar{Y}_{1,t}
= \bar{Y}_{1,t+1} - (\bar{Y}_{0,t+1} - \hat{\beta} \bar{Y}_{0,t})  - \hat{\beta} \bar{Y}_{1,t}
= (\bar{Y}_{1,t+1} - \bar{Y}_{0,t+1} )  - \hat{\beta} ( \bar{Y}_{1,t} - \bar{Y}_{0,t} ).
$$

Second, the coefficient $\hat{\tau}_{\IG}'$ in the ordinary least squares fit
$\hat{\bE}(Y_{t+1}\mid G,Y_t) = \hat{\alpha}  +  \hat{\tau}_{\IG}' G + \hat{\beta}' Y_t $ using all units equals the difference-in-means of $Y_{i, t+1} -  \hat{\alpha} -  \hat{\beta}' Y_{it} $ in the treated and control groups. Therefore,
$$
\hat{\tau}_{\IG}' = (\bar{Y}_{1,t+1} -  \hat{\alpha} -  \hat{\beta}' \bar{Y}_{1,t} ) - (\bar{Y}_{0,t+1} -  \hat{\alpha} -  \hat{\beta}' \bar{Y}_{0,t} )
= (\bar{Y}_{1,t+1} - \bar{Y}_{0,t+1} )  - \hat{\beta}' ( \bar{Y}_{1,t} - \bar{Y}_{0,t} ).
$$

\noindent {\it Proof of Lemma \ref{lemma::diff}.}
The conclusion follows from the law of total probability. We can write $\tilde{\mu}_{0,\IG}-\tilde{\mu}_{0,\DID}$ as
\begin{eqnarray*}
%&&\tilde{\mu}_{0,\IG}-\tilde{\mu}_{0,\DID}  \\
%&=&
&&- \bE(Y_{t+1}\mid G=0) +\bE(Y_{t}\mid G=0) - \bE(Y_{t}\mid G=1) +  \int \bE(Y_{t+1}\mid G=0, Y_t=y) F_{Y_t}(\dd y\mid G=1) \\
&=& -\int \bE(Y_{t+1}\mid G=0, Y_t = y)  F_{Y_t} (\dd y\mid G=0)
+ \int y F_{Y_t} (  \dd y\mid G=0 )
- \int y  F_{Y_t}(  \dd y\mid G=1 ) \\
&&+ \int \bE(Y_{t+1}\mid G=0, Y_t = y)  F_{Y_t} (\dd y\mid G=1) \\
&=& \int \{ \bE(Y_{t+1}\mid G=0, Y_t = y) -y\}  F_{Y_t} (\dd y\mid G=1)
- \int \{ \bE(Y_{t+1}\mid G=0, Y_t = y) -y\}  F_{Y_t} (\dd y\mid G=0).\\
\end{eqnarray*}

\noindent {\it Proof of Theorem \ref{thm::monoATT}.}
The proof relies on a lemma on stochastic ordering in \cite{shaked2007stochastic}. Specifically, for two random variables $A$ and $B$,
$
\pr(A\leq x) \geq \pr(B\leq x)
$
for all $x$ if and only if
$
E\{ u(A) \} \geq E\{  u(B) \}
$
for all non-increasing functions $u(\cdot) $.

Under Condition \ref{assume::stationary}, we have
$
\partial \Delta(y) / \partial y =  \partial \bE(Y_{t+1}\mid G=0, Y_t = y) / \partial y - 1 < 0,
$
i.e., $\Delta(y)$ is a non-increasing function of $y$. Therefore, combining Lemma \ref{lemma::diff}, Condition \ref{assume::monotone}(1) implies $\tilde{\mu}_{0,\DID} \leq   \tilde{\mu}_{0,\IG}$, and Condition \ref{assume::monotone}(2) implies $\tilde{\mu}_{0,\DID} \geq    \tilde{\mu}_{0,\IG}$.

%\begin{proof}[\it Proof of Corollary \ref{thm::IPW}]
%Using the law of iterative expectation, we re-write the definition of $\tilde{\mu}_{0,\IG}$ as
%\begin{eqnarray*}
%\bE\{   \bE(Y_{t+1}\mid G=0, Y_t)  \mid G=1  \}
%&=&\bE\{   G \cdot E(Y_{t+1}\mid G=0, Y_t)    \} / \pr(G=1) \\
%&=&\bE\left\{  e(Y_t) E( Y_{t+1}  \mid G=0,Y_t )  \right\}/ \pr(G=1).
%\end{eqnarray*}
%We can  also re-write $\tilde{\mu}_{0,\IG} ' $ as
%$
%E\left\{  e(Y_t) E( Y_{t+1}  \mid G=0,Y_t )  \right\}/ \pr(G=1).
%$
%Therefore, $\tilde{\mu}_{0,\IG} ' = \tilde{\mu}_{0,\IG}$. Similarly, we can show that $\tilde{\mu}_{0,\DID} ' = \tilde{\mu}_{0,\DID}.$ Therefore, Theorem \ref{thm::monoATT} holds for the weighting estimators.
%\end{proof}

%\bibliographystyle{biometrika}
\bibliographystyle{apalike}
\bibliography{DIDLDV}
\end{document}